\documentclass[aps,amsmath,floatfix,superscriptaddress,prl,reprint]{revtex4-1} 
\usepackage{color} 
\usepackage{refstyle}
\usepackage{amsmath} \usepackage{amssymb} \usepackage{graphicx}
\usepackage[unicode=true,pdfusetitle, bookmarks=true,bookmarksnumbered=false,bookmarksopen=false,
breaklinks=false,pdfborder={0 0 1},backref=false,colorlinks=true] {hyperref}

\newcommand{\revision}[1]{\textcolor{red}{#1}}

\begin{document}

\title{Measuring cohesion between macromolecular filaments, one pair at a time:\\ Depletion-induced
microtubule bundling } 
\author{Feodor Hilitski}
\affiliation{Department of Physics, Brandeis University,
Waltham, MA 02454}

\author{Andrew R. Ward} \affiliation{Department of Physics, Brandeis University, Waltham, MA 02454}

\author{Luis Cajamarca} \affiliation{Department of Physics, University of Massachusetts at Amherst,
Amherst, MA 01003}
 \author{Michael F. Hagan} \affiliation{Department of Physics, Brandeis University,
Waltham, MA 02454} 
\author{Gregory M. Grason} \affiliation{Department of Polymer Science and
Engineering, University of Massachusetts at Amherst, Amherst, MA 01003}
 \author{Zvonimir Dogic} \affiliation{Department of Physics, Brandeis University, Waltham, MA 02454} \email{dogic@brandeis.edu}

\begin{abstract}In presence of non-adsorbing polymers, colloidal particles experience ubiquitous attractive interactions induced by depletion forces. Here, we measure the depletion interaction between a pair of microtubule filaments using a method that combines single filament imaging with optical trapping. By quantifying the dependence of filament cohesion on both polymer concentration and solution ionic strength, we demonstrate that the minimal model of depletion, based on the Asakura-Oosawa theory, fails to quantitatively describe the experimental data. By measuring the cohesion strength in two- and three- filament bundles we verify pairwise additivity of depletion interactions for the specific experimental conditions. The described experimental technique
can be used to measure pairwise interactions between various biological or synthetic filaments and complements information extracted from bulk osmotic stress experiments.\end{abstract}

\maketitle
 Ranging from elastic nanopillar arrays \cite{1} to ropes of carbon nanotubes \cite{2} to
dense chromatin structures \cite{3}, numerous materials of synthetic or biological origin are assembled
from filamentous building blocks. The macroscopic properties of such filamentous materials are governed
not only by the mechanical properties of the constituents, but also by the interactions between them.
These interactions are traditionally measured using bulk osmotic stress experiments in which one
applies an external pressure of known magnitude while simultaneously measuring the filament spacing
using X-ray scattering \cite{4,5,6}. Here we describe a complementary single-filament technique that
directly measures cohesive interactions between a pair of filamentous macromolecules. This approach allows us to assemble bundles in a controlled fashion, with predetermined filament number and binding geometry, yielding information that cannot be accessed by bulk methods. It extends microscopy-based methods developed for measurement of interactions between isotropic, colloidal particles \cite{7,8} to the case of extreme particle anisotropy (e.g. macromolecular filaments).
\begin{figure} \includegraphics[width=0.95\columnwidth]{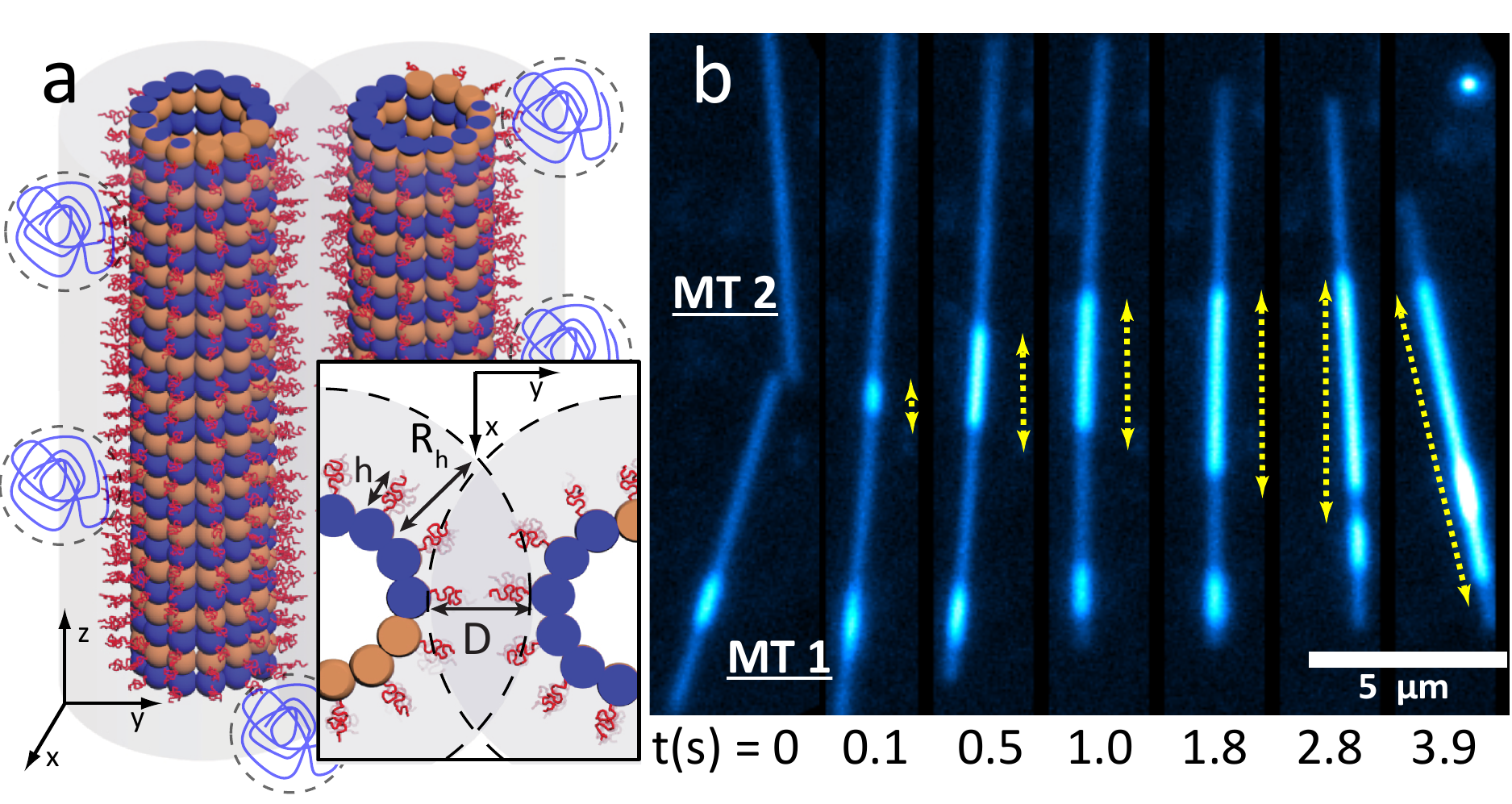}\caption{\label{fig:1}Formation of a microtubule (MT) bundle in quasi-2D chamber. (a) Schematic depiction of the PEG-induced MT bundle, showing depleting polymer, excluded volume shells inaccessible to depleting coils and disordered poly-aminoacid c-terminus tails of tubulin monomers. Inset: Relevant length scales of the bristle-stabilized model: $R_h$, effective radius of the polymer coils; h, bristle height; D, surface-to-surface MT separation. (b) Sequence of \revision{darkfield} microscopy images showing \revision{evolution of a bundle (yellow arrows) formed by two filaments (MT1 and MT2). Short ($\sim 1 \mu m$) fragment bound to MT1 exhibits 1D diffusion along the filament.}} \end{figure}

We study cohesive interactions between microtubules (MTs), cytoskeletal filaments that are assembled
from tubulin heterodimers to form rigid tubular structures with an outer diameter of 25 nm, a contour
length that can reach tens of microns, and a persistence length of a few millimeters \cite{9}. MTs
carry significant negative charge at physiological pH \cite{11}. To assemble bundles, MT filaments were placed in a suspension of non-adsorbing polymers, which induce attractive interactions by the depletion mechanism. \revision{We used either Poly(ethylene glycol) (PEG, MW=20 kDa) or Dextran (MW=500 kDa) as a depletant. The radius of
gyration ($R_g$) of 20k PEG is $\sim$7 nm \cite{15}, Dextran 500k has $R_g\approx 18$ nm \cite{16}.
The depleting polymers were in the dilute regime for all measurements.} As two MTs form a bundle, an additional free volume becomes available to polymer
coils, thus increasing the overall system entropy and resulting in an effective attraction between the
rods (Fig. \ref{fig:1}(a)). Its strength and range can be tuned by changing the polymer concentration and size, respectively \cite{12}. Bundle formation requires high ionic strength in order to screen the repulsive electrostatic interactions between the filaments \cite{13}.

We directly visualized multi-step formation of bundled MTs in the presence of the depleting agent (Fig. \ref{fig:1}(b)). First, two diffusing MTs encountered each other and formed a bundle with a random overlap. Subsequently, the bundle maximized the MT overlap. To understand such a process, consider the free energy of a bundle given by $G =-\lambda L$, where $\lambda$ is the cohesive free energy per unit length and $L$ is the filament overlap length. The magnitude of the retraction force is given by $f=-\partial G/\partial L=\lambda$. Therefore, measuring the retraction force directly quantifies the depletion interaction as characterized by the cohesion energy per unit length, $\lambda$.

\begin{figure} \includegraphics[width=1\columnwidth]{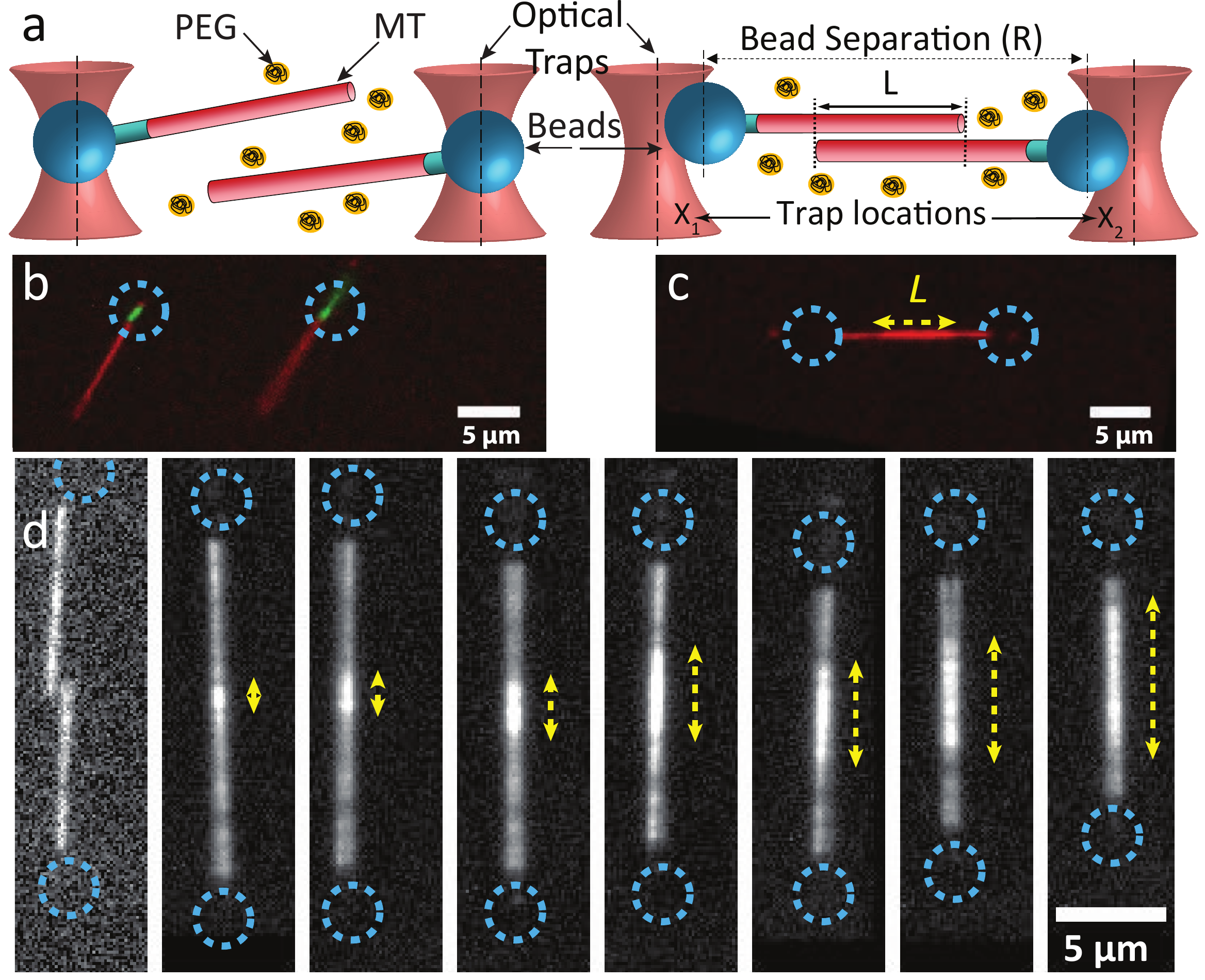}\caption{\label{fig:2}Experimental setup used to measure MT cohesion strength. (a) Left: two unbundled optically trapped bead-MT complexes. Right: upon bundle formation the depletion force pulls beads closer together, away from the trap centers. (b) Composite fluorescence image of the MT filaments showing biotin-free elongated segments (red) and biotin labeled seeds (green). Beads attached to seeds are not visible. Optical traps are indicated by the dashed lines. (c) Composite fluorescence image showing MTs from panel (b) forming a bundle, visible as the bright red region. Bundle overlap region and trap locations are indicated by the yellow and blue dashed lines, respectively. (d) Bundle overlap length (yellow arrows) is controlled by the independent movement of two optical traps (blue dashed lines).} \end{figure}

The experimental system used to directly measure $G$ and determine $\lambda$ is illustrated in Fig.
\ref{fig:2}. Briefly, we prepared segmented MTs in which only a short portion (seed) is labeled with
biotin. The biotinylated seeds and the
biotin-free elongated segments of MTs were labeled with different fluorescent dyes \footnote{See Supplemental Material for materials and methods}. The entire MTs were
stabilized with GMPCPP, a non-hydrolysable analog of GTP, thus ensuring that there is no dynamic
instability. Optically trapped Neutravidin coated silica beads were manually attached to biotin labeled MT seeds (Fig. \ref{fig:2}(a, b)). Subsequently, the two MTs, each attached to a single bead, were brought into close proximity \revision{to facilitate bundling} \footnote{See Supplemental Material for Supplemental Videos 1,2}. Once the MT bundle formed (Fig. \ref{fig:2}(c)), we verified that the MTs overlapped only along the biotin-free elongated segments. This ensured that Neutravidin, which desorbs from beads, did not crosslink biotinylated parts of both filaments. Our experimental setup allowed for control of the MT overlap length with nanometer accuracy (Fig. \ref{fig:2}(d)).

\begin{figure} \includegraphics[width=1\columnwidth]{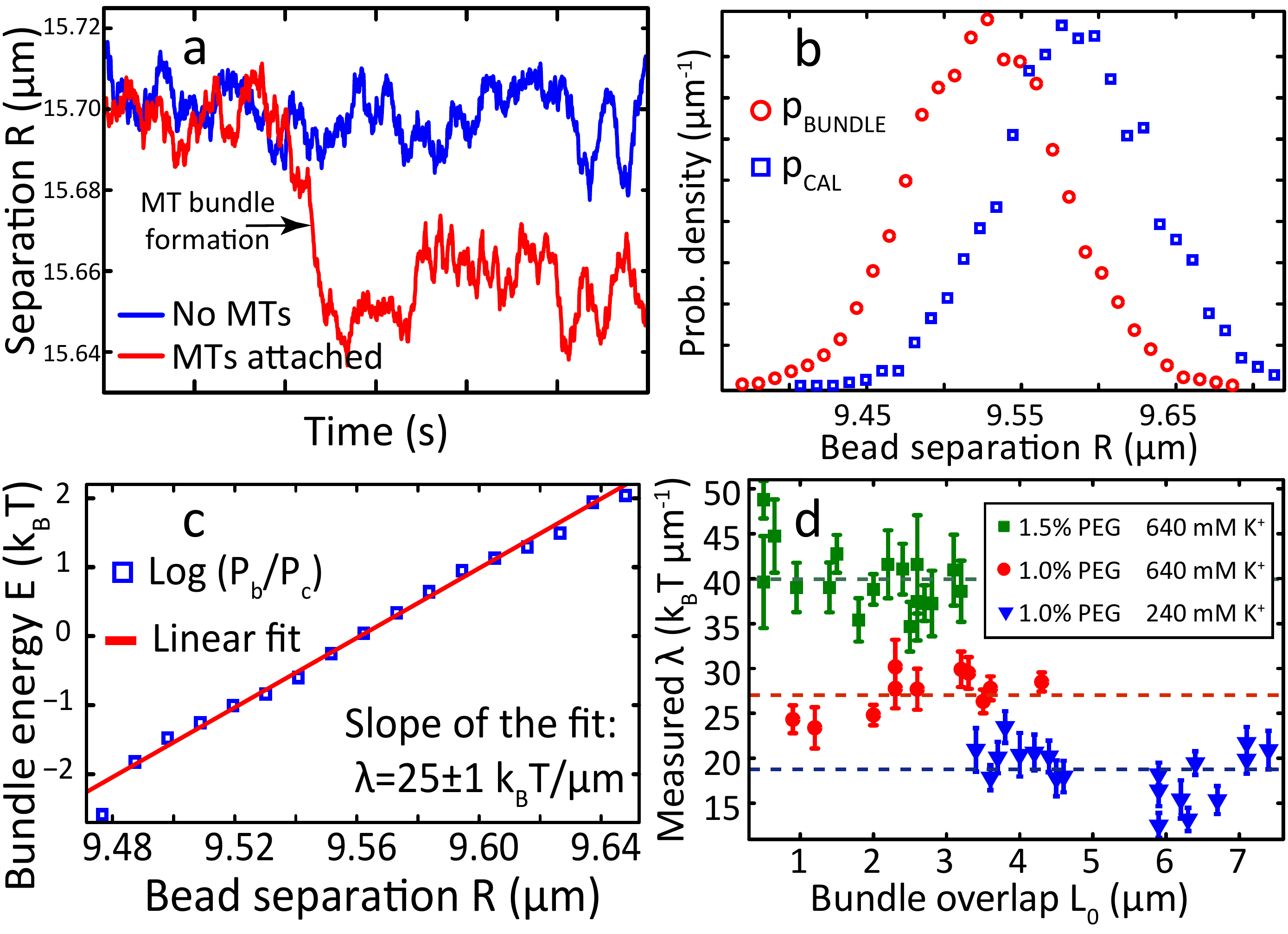}\caption{\label{fig:3}Analysis of the experimental data. (a) Fluctuations in the separation of two beads ($R$) with and without bundled MTs (blue line). MT bundle formation is accompanied by a rapid reduction of $R$ (red line). (b) Probability distributions $P_{bundle}(R)$ and $P_{calibration}(R)$ are obtained by sampling statistically independent configurations of the two beads. (c) The bundle free energy, $G$, scales linearly with the bead separation. Slope of the weighted linear fit gives the value
of the cohesion strength, $\lambda$. (d) $\lambda$ is independent of the overlap bundle $L$. Dashed
lines correspond to the mean cohesion strength. Error bars for the individual measurements represent 95\%
confidence intervals.} \end{figure}

Before bundle formation, each bead with an attached MT fluctuated around the minimum of the harmonic
potential imposed by the optical traps. Upon bundle formation, the average bead separation, $R$,
decreased due to the retraction force (Fig. \ref{fig:3}(a)). While measurement of the reduction in
average spacing for known trap stiffness directly yields the magnitude of the retraction force $f =
\lambda$, we improved statistics by incorporating the fluctuations around the equilibrium position.
Specifically, we  experimentally implemented a form of umbrella sampling to measure the free energy, $G$, of the bundle as a function of $R$ \cite{17,18}. As beads fluctuated in the optical traps \footnote{Supplemental Video 3},
we measured their center-to-center separation $R$ and constructed normalized probability distributions:
$P_{bundle}(R)$ and $P_{calibration}(R)$ for the bundled and unbundled configurations, respectively
(Fig. \ref{fig:3}(b)). Both distributions had Gaussian shapes of equal width, which was determined by the
stiffness of the optical traps alone. Presence of the bundle reduced the mean bead separation. Since the separation
$R$ and the overlap length $L$ are simply related \footnote{$R=L_{MT1}+L_{MT2}-L$, where $L_{MT1}$ and $L_{MT2}$ are the length of the bundled filaments}, the bundle
free energy as a function of the overlap length, $L$, is given by:
\begin{equation} G(L)=-k_{B}T\
\log\left(\dfrac{P_{bundle}(L)}{P_{calibration}(L)}\right)+const \label{eq:1} \end{equation}
We extracted $G(L)$ by using the measured distributions in Eq. \ref{eq:1}. The cohesion energy per unit length is given by $\lambda=-\partial G / \partial L$, which was calculated from the slope of the weighted linear fit to the experimental points (Fig. \ref{fig:3}(c)). For example, the
probability distributions depicted in Fig. \ref{fig:3}(b) yielded $\lambda = 25 \pm 1$ $k_B$T/$\mu$m,
which is equal to the retraction force of $\sim$0.1 pN. A single experiment samples only a small region
of the bundle overlap $(\approx100-150$ nm) that is accessible by thermal fluctuations. To extend
this range, we manually changed the overlap length and repeated the measurement. We found $\lambda$ to
be independent of the MT bundle overlap, $L$, in agreement with our initial assumptions (Fig.
\ref{fig:3}(d)).

\begin{figure} \includegraphics[width=0.8\columnwidth]{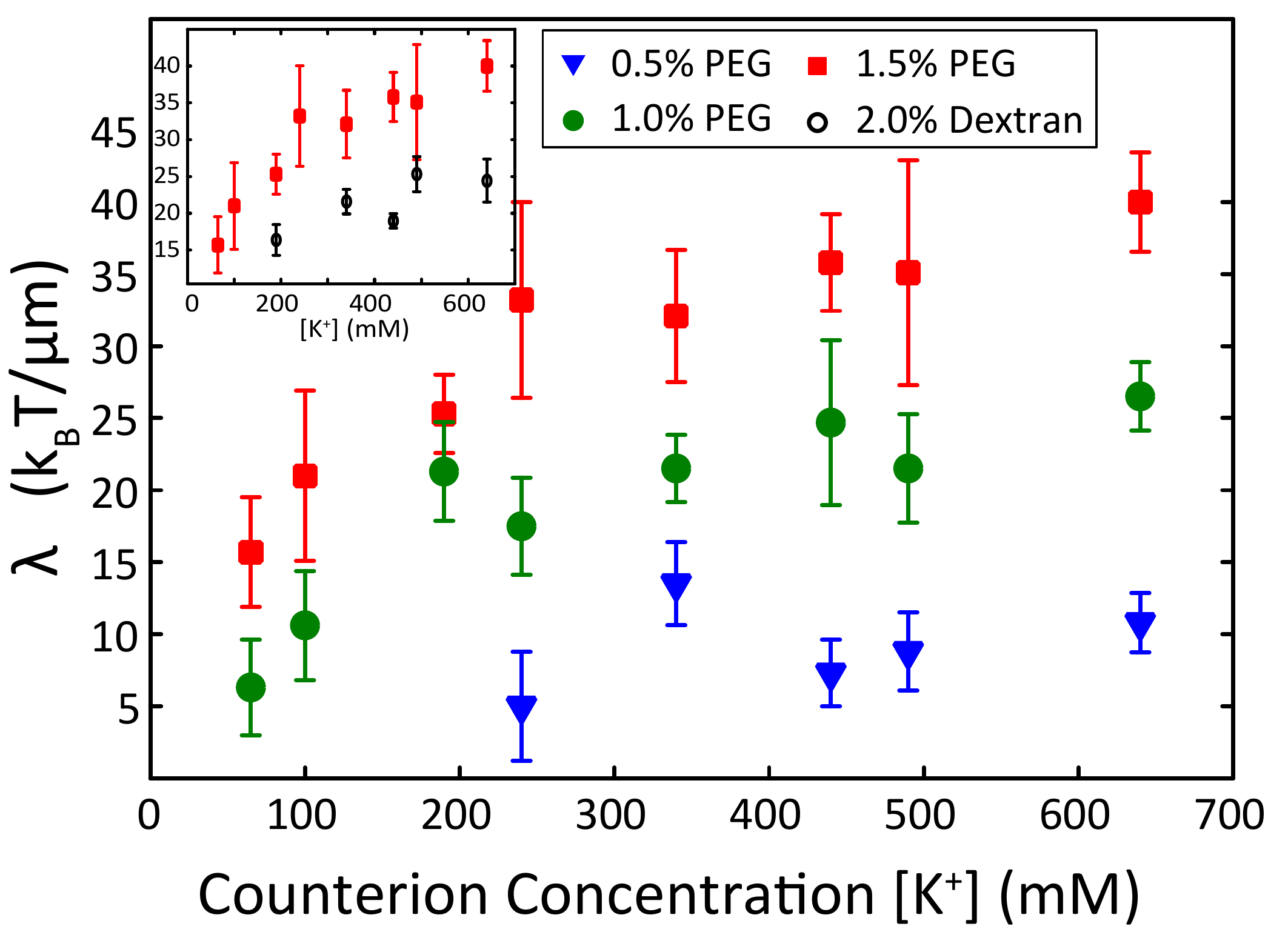}\caption{\label{fig:4}Dependence of $\lambda$ on the counterion concentration [K$^+$] for several PEG concentrations. Inset: Dependence of $\lambda$ on [K$^+$] for 2.0\% Dextran and 1.5\% PEG. Each point is an average of $N=5-60$ independent measurements. Error bars represent standard deviation.} \end{figure}

\begin{figure} \includegraphics[width=0.75\columnwidth]{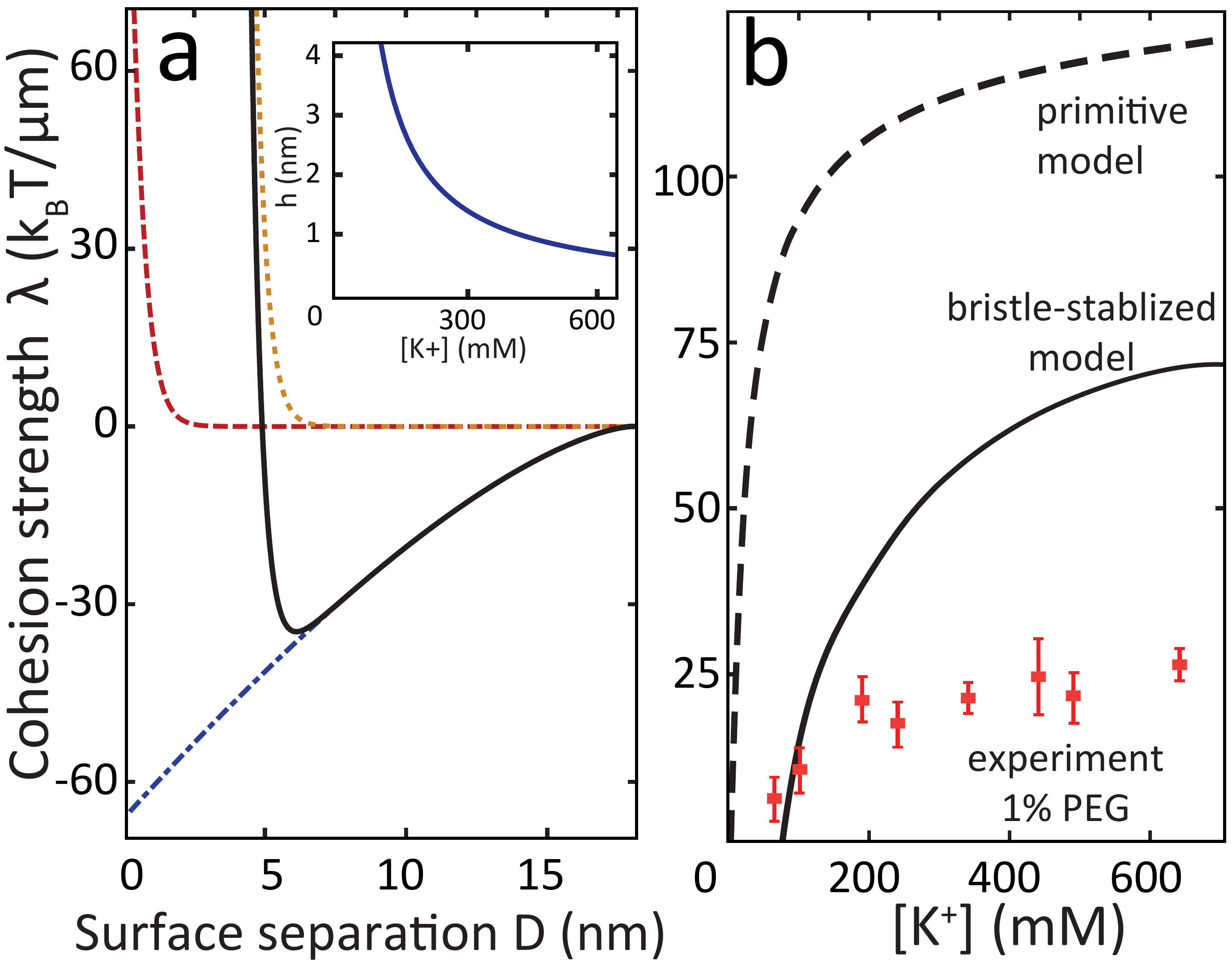}
\caption{\label{fig:5}Comparisons of theoretical models of the cohesion strength. (a) Effective interactions between two
parallel MTs (solid black curve) as a function of filament surface separation, $D$. The effective
potential is composed of the depletion attraction (dash-dotted blue curve) and repulsive potential due
to charged hollow cylinders (red dashed curve) and charged bristles (dotted light brown curve). Inset:
The predicted bristle height $h$ vs. salt concentration. (b) Theoretical predictions of the cohesion strength for the primitive model (dashed curve), the bristle-stabilized
model (solid curve) and experimental results (squares) for PEG 1.0\%.} \end{figure}

We measured how the cohesion strength depends on both the suspension ionic strength and depleting
polymer concentration. For a fixed ionic strength, $\lambda$ increased linearly with increasing polymer
concentration. For a fixed polymer concentration, $\lambda$ decreased with decreasing ionic strength
(Fig. \ref{fig:4}). These results can be qualitatively understood by considering that the cohesive
potential between aligned MTs is governed by a combination of the attractive depletion interaction and
electrostatic repulsion (Fig. \ref{fig:5}(a)). In the Asakura-Oosawa (AO) model, the depletion strength
depends linearly on the depletant concentration, in agreement with the experimental findings. Upon
decreasing ionic strength, the range of the electrostatic repulsion increases,
thus leading to decreased cohesion strength as observed in the experiments. When the ionic strength is
sufficiently low, the depletion attraction is unable to overcome repulsion between the negatively
charged filaments, thus suppressing bundling.

To quantitatively assess the relationship between measured MT cohesion and physical and structural
features of MTs, we developed two theoretical models. In the simplest \textit{primitive} model, MTs
were treated as uniformly-charged, hollow cylinders of radius $a=12.5$ nm with surface charge
density $\sigma_s$= -0.23\ e/nm$^2$ (-23e per tubulin dimer) \cite{11}. Electrostatic repulsion was
computed via the linearized Poisson-Boltzmann theory, and the depletion-induced attraction was computed
by modeling polymer coils as effective hard spheres with radius $R_h = 2R_g / \sqrt{\pi}$ \cite{19}.
The binding free energy predicted by the minimal-energy separation of the primitive model overestimates
the measured cohesion strength by a factor of $\sim5-7$, implying an overestimation of depletion
attraction and/or an underestimation of repulsions. To explore the possibility that repulsive forces
generated by negatively charged, flexible "bristle-like'' c-terminal tails of tubulin \cite{20,21}
contribute to this discrepancy, we developed a \textit{bristle-stabilized} MT model, which includes
additional salt-dependent electrostatic forces between c-terminal bristles anchored to MT surfaces (Fig
\ref{fig:1}(a, inset)). Salt-dependent longer-range repulsive forces generated by the c-terminus
bristles have been implicated in previous experimental measurements of inter-MT spacing and tubulin
interactions \cite{22,23}. Here, we adopted the electrostatic brush model of Pincus \textit{et al.}
\cite{24} to calculate the dependence of bristle height, $h$, on the ionic strength. Treating the
bristles as 13-segment chains \cite{25} (segment size 0.5 nm) with charge $Q = -8$e per bristle as (an
approximately) flat brush of areal density of one bristle per tubulin monomer $\sigma_B= 1/4^2\
 $nm$^{-2}$, we predicted bristle heights to vary from 2.6 nm to 4.8 nm over the range of salt
concentrations measured, scaling roughly as $h\sim [K^+]^{-1/3}$ in the high-salt limit (Fig.
\ref{fig:5}(a, inset)). We assumed uniformly stretched bristles with the bristle charge density at
radial distance $r$ from the MT center given by $\rho_B (r) = Qa\sigma_B / rh$. Electrostatic
repulsion in the bristle-stabilized model was computed from the linearized-PB interaction between
superposed “bare” surface charge of MTs and bristle charge density, $\rho_f(r) = \sigma_s
\delta(r-a)+\rho_B(r)$, while depletion was treated as before, such that polymers were assumed to
penetrate the bristles freely. These interactions increase filament separation by roughly twice the
bristle height at the expense of decreasing depletion attractions. While calculated binding energies
were closer to the experimental values, they still overestimated the measured $\lambda$ by a factor of
$\sim$2-3 (Fig. \ref{fig:5}(b)). Comparing the \textit{primitive} and \textit{bristle-stabilized} model
predictions suggests that the larger surface separation due to charged c-terminus
bristles is fundamental to MT-MT interactions. However, in light of apparent overestimation of the
polymer-induced depletion by the AO theory, a more detailed representation of polymer interactions with
the MT geometry is needed to quantitatively recapitulate measured cohesion strengths.

\begin{figure}
\includegraphics[width=0.85\columnwidth]{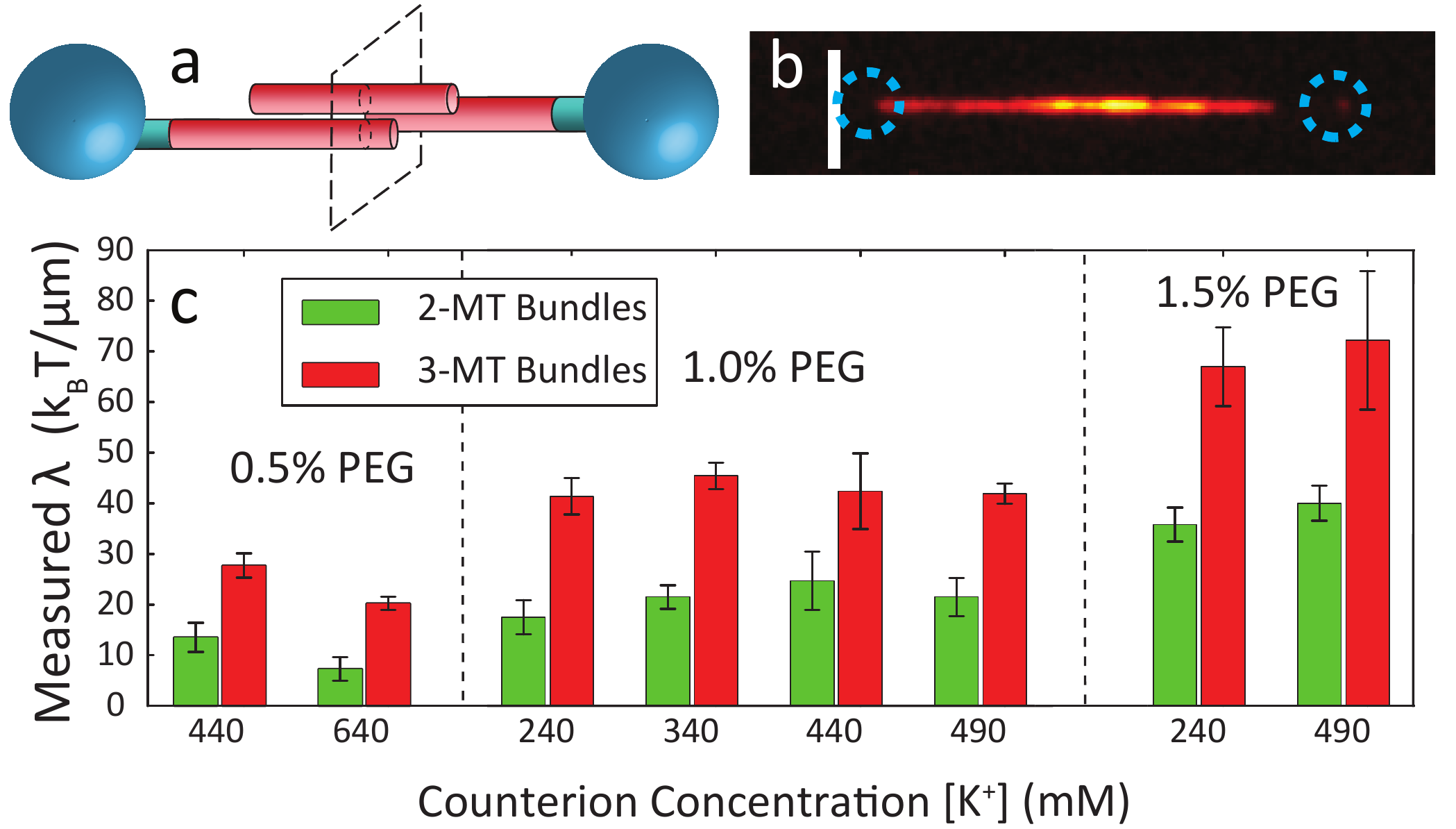}\caption{\label{fig:6}Pairwise
additivity of the depletion interaction. (a) Schematic representation and (b) fluorescence image of a
three-filament bundle.The third MT is not attached to the beads.  Scale bar, 5 $\mu$m. (c) For a wide range of experimental
conditions, $\lambda$ for three-MT bundles is double the value for the two-MT bundles. Error bars represent standard deviation.}
 \end{figure}

Depletion interactions between spherical colloids and polymers are pairwise additive as long as the
depleting polymer is significantly smaller than the spherical colloid \cite{26,27}. To investigate the
additivity of the depletion between rod-like particles, we characterized properties of multi-filament
bundles. We induced the formation of a three-MT bundle by bringing a third filament near an existing
two-MT bundle. Once adsorbed, the third MT always migrated to the existing two-filament overlap region,
since such configurations minimize the depletant excluded volume (Fig. \ref{fig:6}(a, b)). The cohesion strength, $\lambda$, for the 3-MT bundles was consistently twice the value measured for the 2-MT bundles (Fig. \ref{fig:6}(c)). These experiments indicate that depletion interactions with a
$\sim$7 nm range are still pairwise additive for the filaments with hardcore diameters of $\sim$25 nm.
The electrostatic interactions decrease the range of the depletion attraction and thus likely further
reduce the importance of many-body overlap configurations.

Polymer-induced depletion interactions with nanometer range have been extensively studied between
micron-sized colloidal particles \cite{8,28,29}. Here we have measured the depletion interactions in
the protein limit \cite{30,31}, in which the size of the non-adsorbing polymer ($R_g$) approaches the
size of the colloidal object (MT filament diameter). We found significant discrepancies between our
measurements and predictions of a simplified AO model. These were only partially reconciled by a more
refined microscopic description of MT structure. To ultimately discern the source of this discrepancy,
several other effects, beyond the scope of this manuscript, need to be investigated. For example, the
AO model is only accurate when polymer coils are significantly smaller than colloidal particles
\cite{30}. Though the accuracy of the AO model has been studied for spherically symmetric particles,
there are few equivalent studies for rod-like colloids \cite{32}. Furthermore, bending degrees of
freedom are severely constrained in a bundle, corresponding to an additional entropic cost of bundling
for finite-persistence length filaments. This entropy loss was important for understanding
depletion-induced bundling of actin filaments \cite{33}; since the bending entropy loss scales
inversely with the fourth root of the bending stiffness, it could yield significant corrections even
for stiff MTs. Finally, we have treated the depletant as ideal non-adsorbing polymers; excluded volume
interactions could affect the magnitude of the depletion attraction \cite{34}.

 \revision{In general, depletion-induced filamentous bundles can exist in two distinct states - dynamical, where the sliding friction of constituent filaments is governed by hydrodynamic interactions, and static, where solid-like friction dominates \cite{Ward}.Our technique measures cohesion strength for dynamical bundles, whose filaments are freely sliding with respect to each other and thus minimize the free energy on experimental timescales. For MT filaments, transition between the two regimes is governed by the concentrations of both depletant and counterions as well as by some structural modifications of tubulin (for example, removal of the c-terminus brush) \cite{Ward}.}

In conclusion, we have developed a technique to measure attractive depletion interactions between
filamentous structures, and applied it to reveal intriguing properties of the depletion-driven MT bundles. Most materials obey Hooke's law, so that the deformation energy scales with square of the applied strain. In comparison, the free energy of MT bundles scales linearly with the applied strain, indicating that such assemblages behave as a constant force transducer with effectively zero-stiffness. Our results are relevant for biological or synthetic systems where depletion forces due to globular polymers or other non-adsorbing particles play an important role \cite{35,36}. They are also important in materials science, where the depletion effect is used to engineer interactions between colloids and to drive self-assembly \cite{32,37,38,39,40,41,42,43}. \revision{The methods described here can be used in studies of MT-associated motor enzymes and cross-linking proteins (MAPs) \cite{Bieling2010,Forth2014,Lansky2015}} and extended to other filamentous systems and alternative interactions such as counterion-induced bundling of polyelectrolyte filaments \cite{44,48,45,46,47}.

 \begin{acknowledgments} FH and ZD were supported by Department of Energy, Office of Basic Energy Sciences under Award DE-SC0010432TDD. AW acquired preliminary data with support of NSF-MRI-0923057. MFH developed the theoretical model with support of NSF-CMMI-1068566 and NSF-MCB-0820492. Theoretical work of LC and GG was supported by NSF-CMMI-1068852. We also acknowledge use of MRSEC optical microscopy facility which is supported by NSF-MRSEC-0820492.
\end{acknowledgments}

\bibliography{references} 
\end{document}